\documentclass[sigconf,screen]{acmart}

\setcitestyle{nocompress} 

\AtBeginDocument{%
  
}

\copyrightyear{2026}
\acmConference[Poster at SecDev '26]{Poster at ACM Secure Development Conference}{July 5--6, 2026}{Montreal, QC, Canada}
\acmBooktitle{Poster at ACM Secure Development Conference (SecDev '26), July 5--9, 2026, Montreal, QC, Canada}
\setcopyright{cc}
\setcctype{by}

\begin{document}


\title{Evaluating Retrieval-Augmented Generation for Explainable Malware Analysis}

\author{Jayson Ng}
\affiliation{
  \institution{New York Institute of Technology}
  \city{Vancouver}
  \state{BC}
  \country{Canada}
}
\email{jng10@nyit.edu}

\author{Amin Milani Fard}
\affiliation{
  \institution{New York Institute of Technology}
  \city{Vancouver}
  \state{BC}
  \country{Canada}
}
\email{amilanif@nyit.edu}

\begin{abstract}
Large Language Models (LLMs) are increasingly being used as security engineering tools to summarize and explain malware behavior to analysts. A common assumption is that Retrieval-Augmented Generation (RAG) improves explanation quality by injecting external security knowledge. In this work, we empirically evaluate this assumption for malware explanation using VirusTotal reports as structured input. Across multiple LLMs, we find that RAG frequently degrades explanation quality by introducing distracting or weakly related context and adding narrative noise or generic write‑ups. Our results highlight a practical risk in security-critical pipelines for malware explanation that RAG can be counterproductive when structured security evidence is already sufficient. We argue that malware explanation is primarily a signal-extraction task, not a knowledge-retrieval problem, and outline design recommendations for secure development workflows.
\end{abstract}

\keywords{Malware Analysis, Secure Development, Explainable AI, LLM, RAG}

\maketitle

\section{Introduction}

Understanding why a binary is malicious is as important as determining whether it is malicious. While platforms such as VirusTotal provide rich detection labels, behavioral traces, and metadata, they do not offer analyst‑ready explanations that connect low‑level indicators to high‑level attacker intent. These outputs lack a coherent causal narrative explaining why certain behaviors matter, how execution unfolds, or how signals indicate malicious activity. Such explainability is essential for threat hunting, incident response, and regulatory or audit requirements.

Recently, LLMs have been proposed as security engineering tools to translate low-level indicators (e.g., API calls, registry edits, and network activity) into natural-language explanations that support triage, incident response, and reporting. For example, MalGPT \cite{saqib2025malgpt} demonstrates that generative models can produce meaningful explanations from malware binaries by learning latent behavioral patterns. RAG \cite{lewis2020retrieval} is often treated as a default enhancement for LLM-based systems, and has been applied to related pipelines such as CVE association and reasoning over decompiled artifacts \cite{cristea2026malcve,pnnl2024cyrag}. VulRAG \cite{du2024vul} shows that incorporating structured domain knowledge improves reasoning about software vulnerabilities and reduces hallucinations. However, prior studies do not evaluate whether RAG improves explanation quality in malware analysis. Unlike domains that rely on external documentation, malware reports often encapsulate sufficient contextual evidence, raising the operational risk that indiscriminate retrieval introduces distraction or noise leading to misleading explanations in security workflows. 

\section{Proposed Approach}

We evaluate the explainability of LLMs with and without RAG for malware explanation tasks when structured evidence already exists. 

\begin{figure}
\centerline{\includegraphics[trim = 10mm 5mm 15mm 2mm,scale=0.5]{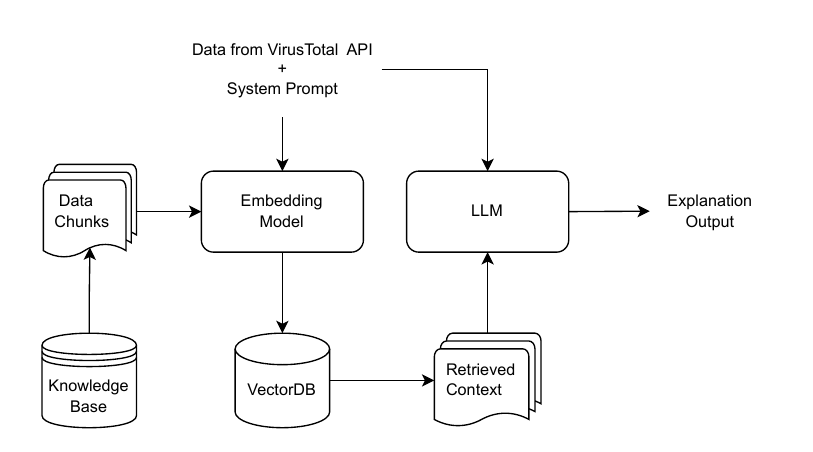}}
\caption{RAG LLM architecture for malware explanation.}
\label{RAGLLM}
\end{figure}

\textbf{Implementation.} We use LlamaIndex as the RAG orchestration framework, ChromaDB for persistent vector storage, and OpenRouter as the LLM gateway. We apply all‑MiniLM‑L6‑v2 and OpenAI's text‑embedding‑3‑large for embeddings. Documents are chunked into 1024 tokens with a 100‑token overlap. Retrieval employs top‑$k$ similarity search with a 0.5 threshold. The knowledge base consists of 26 documents parsed using LlamaIndex's SimpleDirectoryReader. Our dataset is derived from the MalGPT corpus \cite{saqib2025malgpt} with 1,702 VirusTotal reports, including structured indicators such as API calls and registry operations. MD5 hashes serve as identifiers. Our implementation is available for download\footnote{\url{https://github.com/nyit-vancouver/RAG-Explain-Malware}}.

\textbf{System prompt.} The prompt has 10 sections: (1) one‑line verdict with confidence, (2) non‑technical summary, (3) evidence bullets with JSON citations, (4) malware family mapping, (5) grouped Indicator of Compromises (IoCs) with paths, (6) behavioral summary, (7) confidence score, (8) recommended actions for technical and non‑technical audiences, (9) optional FAQ, and (10) raw evidence appendix. This structure enforces consistency and traceability.

\textbf{Knowledge base.} The knowledge base integrates four components: (1) the MITRE ATT\&CK framework (Enterprise, ICS, Mobile); (2) research and best‑practice guidance, including MalGPT and CISA/MITRE mapping recommendations; (3) VirusTotal technical documentation covering behavioral schemas, reports, and IoC definitions; and (4) malware intelligence datasets, including malware family taxonomies and CVE records.

\textbf{Evaluation.} We conduct an ablation study comparing LLMs with and without RAG against GPT‑5.1 (Table~\ref{tab:results}). Each model runs under identical prompts using either VirusTotal JSON alone or JSON augmented with retrieved context. To isolate retrieval effects, no fine‑tuning is performed. We do not compare directly with MalGPT, as it represents a different paradigm based on a trained multi‑modal architecture rather than external retrieval. Since our focus is on retrieval‑induced effects when explaining structured sandbox evidence, such a comparison would not isolate the impact of RAG. We measure explanation quality using BERTScore \cite{Zhang2020BERTScoreICLR}, which captures semantic similarity using contextual embeddings and is robust to paraphrasing and structural variation suitable for evaluating malware explanations where equivalent meanings may be expressed using different terminology. It is also sensitive to semantic drift introduced by loosely related retrieved passages.

\begin{table}[t]
    \footnotesize
    \begin{tabular}{llll}
    \toprule
    \textbf{Model} & \textbf{\# Parameters} & \textbf{Embeddings} & \textbf{BERTScore} \\
    \midrule
    GPT OSS & 21B & None & 0.8407 \\
    GPT OSS & 21B & all‑MiniLM‑L6‑v2 & 0.8184 \\
    GPT OSS & 21B & text‑embedding‑3‑large & 0.8493 \\
    DeepSeek-R1 & 671B & None & \textbf{0.8617} \\
    DeepSeek-R1 & 671B & all‑MiniLM‑L6‑v2 & 0.8583 \\
    DeepSeek-R1 & 671B & text‑embedding‑3‑large & 0.8582 \\
    GPT‑5.1 & est.\ 2.5T & None & \textbf{1.0000} \\
    GPT‑5.1 & est.\ 2.5T & all‑MiniLM‑L6‑v2 & 0.8778 \\
    GPT‑5.1 & est.\ 2.5T & text‑embedding‑3‑large & \textbf{0.9047} \\
    \bottomrule
    \end{tabular}
    \caption{Malware explanation compared to GPT‑5.1 w/o RAG.}
    \label{tab:results}
\end{table}

\section{Discussion}

\textbf{Results analysis.} Based on commonly observed ranges in the literature, BERTScores 0.9-1 indicate excellent alignment, scores between 0.85 and 0.9 reflect strong semantic similarity, and scores between 0.8 and 0.85 indicate moderate quality. Differences as small as 0.01–0.02 are meaningful in technical domains. Excluding GPT‑5.1, the highest score is achieved by DeepSeek-R1 without RAG (0.8617), indicating that retrieval does not effectively improve explanation quality. 
As shown in Table~\ref{tab:results}, the all‑MiniLM‑L6‑v2 retriever degrades performance for both models, particularly GPT‑OSS‑20B. While text‑embedding‑3‑large improves GPT‑OSS‑20B slightly, it does not benefit DeepSeek-R1. Compact embeddings often retrieve semantically similar but low‑utility passages, such as generic malware descriptions, which distract the generator. Prior work formalizes this distracting effect and shows that weakly relevant retrieval degrades performance \cite{Amiraz2025Distracting}. Long‑context RAG studies similarly observe performance decline as retrieved context accumulates hard negatives \cite{Jin2025LongContextRAG}. These findings align with our observations. VirusTotal JSON reports already provide tightly scoped behavioral evidence. Injecting external descriptions shifts model attention toward plausible but irrelevant narratives, reducing semantic alignment. For strong models such as DeepSeek-R1, additional context appears redundant or mildly contradictory \cite{joren2025sufficient}. In this setting, malware explainability is primarily a signal‑extraction problem rather than a knowledge‑retrieval one. Retrieval thresholds were intentionally permissive to expose potential distractions.

\textbf{Security engineering implications.} Our findings highlight important implications for secure system design. The risk of automation bias increases when RAG-based explanations that drift from the underlying evidence are not manually verified. Analysts benefit more from consistency and traceability than from expanded narratives. From a tooling perspective, malware explanation is better framed as structured signal extraction rather than knowledge synthesis. In security-critical pipelines, adding context without clear necessity increases cognitive load and the risk of error.

\textbf{Recommendations.} Our results suggest that when base evidence is already sufficient, RAG can degrade explanation quality. 
To mitigate retrieval‑induced noise, we recommend: (1) skipping RAG when structured reports are sufficient using context‑sufficiency predictors \cite{joren2025sufficient}; (2) applying domain‑tuned re‑rankers with strict $k$ limits to avoid hard negatives \cite{Jin2025LongContextRAG}; (3) replacing fixed‑window chunking with behavior‑ or IoC‑centric chunks enriched with metadata \cite{Evidently2025RAGEvalGuide}; (4) adversarial fine‑tuning with hard distractors \cite{Amiraz2025Distracting}; and (5) exploring alternatives such as cache‑augmented generation, structured context engineering, or knowledge‑graph‑based augmentation. 

\section{Conclusion and Future Work}

While RAG is often proposed to enhance LLMs with external knowledge, our experiments show that low‑relevance or poorly retrieved context can degrade malware explanation quality. This aligns with known limitations of RAG and suggests that retrieval is not universally beneficial—particularly in domains driven by structured reasoning rather than knowledge completion. Malware triage using VirusTotal primarily involves extracting signals from structured artifacts such as behaviors, signatures, and hashes. When this evidence is already sufficient, additional retrieved text can dilute salient signals, increase cognitive load, and distract the model. As future work, we will explore alternatives to RAG, including Cache‑Augmented Generation and Context Engineering, which better preserve structured evidence without introducing noise.

\bibliographystyle{ACM-Reference-Format}
\bibliography{references}

\end{document}